\documentclass[sigconf]{acmart}
\AtBeginDocument{%
  }

\copyrightyear{2026}
\acmYear{2026}
\setcopyright{cc}
\setcctype{by}
\acmConference[CCS '26]{Proceedings of the 2026 ACM SIGSAC Conference on Computer and Communications Security}{November 15--19, 2026}{The Hague, Netherlands}
\acmBooktitle{Proceedings of the 2026 ACM SIGSAC Conference on Computer and Communications Security (CCS '26), November 15--19, 2026, The Hague, Netherlands}
\acmDOI{10.1145/3830454.3846451}
\acmISBN{979-8-4007-2871-6/2026/11}
\newcommand{\pw}{\textsc{ProofWeave}}

\begin{document}

\title[Poster: Towards ProofWeave]{Poster: Towards ProofWeave: A Privacy-Minimised, Integrity-Anchored Evidence Plane for Continuous Agentic Assurance}

\author{Guy Lupo}
% \authornote{Corresponding author.}
\affiliation{
  \institution{Swinburne University of Technology}
  \city{Melbourne}
  \state{VIC}
  \country{Australia}
}
\email{glupo@swin.edu.au}

\author{Nguyen Hung Nguyen}
\affiliation{
  \institution{Swinburne University of Technology}
  \city{Melbourne}
  \state{VIC}
  \country{Australia}
}
\email{nguyenhungnguyen@swin.edu.au}

\author{Viet Vo}
\correspondingauthor
\affiliation{
  \institution{Swinburne University of Technology}
  \city{Melbourne}
  \state{VIC}
  \country{Australia}
}
\email{vvo@swin.edu.au}

\author{Chamikara M.A.P.}
\affiliation{
  \institution{CSIRO}
  \city{Melbourne}
  \state{VIC}
  \country{Australia}
}
\email{chamikara.arachchige@csiro.au}

\author{Guangdong Bai}
\affiliation{
  \institution{City University of Hong Kong}
  \country{Hong Kong}
}
\email{g.bai@cityu.edu.hk}

\renewcommand{\shortauthors}{Guy Lupo, Nguyen Hung Nguyen, Viet Vo, Chamikara M.A.P., and Guangdong Bai}
%% No italics, no superscripts, not anonymous
%% Use footnote or author note to identify equal contribution, shared contribution, and/or contact author info


\begin{abstract}
Agentic AI systems increasingly act via tools, memory, delegation, and external services, yet post-hoc observability rarely proves that a policy-relevant action was checked by the intended control under the policy then in force. Assurance may therefore rely on evidence that is incomplete, privacy-leaking, mutable, or detached from governing policy.

We introduce \pw{}, a record-time chain of evidence that binds agent intent or action, control response, and policy snapshot into a privacy-minimised, integrity-anchored transaction. Each transaction is committed to an append-only ledger and materialised into an evidence graph for deterministic validation. In a minimal secret-exfiltration scenario, \pw{} reduces candidate bindings from up to $10{,}201$ to one, validation operations from up to $10{,}201$ to approximately $26$, and estimated evidence storage from $0.79$~MiB to $0.15$~MiB per project.

\end{abstract}

\maketitle

\section{Motivation and Practical Demands}
% Agentic AI systems do not merely produce text. They read files, invoke tools, call APIs, write memory, delegate work, and affect external environments. Their security controls can also be agentic: a detector may use a model to explore code, retain intermediate facts, validate candidate paths, and emit a finding. A safe-looking output or plausible report does not reveal whether a required validator ran, whether execution completed, whether a provider fault truncated the result, or which policy applied. This is an enforcement-attribution problem for output-only and surrogate testing~\cite{surrogatetesting2026}.

Agentic AI systems do not merely produce text: they read files, invoke tools, call APIs, write to memory, delegate tasks, and interact with external services. As a result, inspecting only the final response is no longer enough. A benign-looking answer may reflect (i) a risky action the agent never attempted, (ii) an attempt that was blocked by runtime control, or (iii) an attempt that executed but left little or no visible trace in the final output. Distinguishing these cases is the \emph{enforcement attribution problem}.

Resolving enforcement attribution requires evidence from three sources: the agent's declared purpose and attempted action, the control's observations and decision, and the policy service's record of the rules in force at the time. Together, these sources impose three requirements: capture security-relevant events at action boundaries, prove that the intended control evaluated the action under the applicable policy, and protect the resulting evidence from privacy and integrity risks. Conventional logs retain these perspectives in producer-specific formats. At scale, action events, control decisions, and policy snapshots must therefore be correlated quickly enough for real-time monitoring while remaining rich enough for later investigation. Subsequent joins depend on stable identifiers, clock alignment, retained payloads, source authentication, and trust in the graph reconstructed from those logs. System provenance supports post hoc reconstruction, while provenance-based intrusion detection models provide information-flow and causal relations for backward and forward tracing~\cite{zipperle2022pids,li2021systemprovenance}. Agent-native audit graphs, e.g., Agent-BOM, similarly represent the model, tool, memory, capability, semantic state, and cross-agent activity~\cite{agentbom2026}. Trust observability, however, poses a narrower record-time question: not only \emph{what happened}, but whether the relevant control ran under the applicable policy when the action was attempted. The missing property is therefore a common evidence protocol that preserves source identity, policy-at-time, privacy, and integrity while supporting controls and analyses that were not hard-coded into the original detector.

% \vspace{-5pt}
\paragraph{Four evidence-plane gaps.}
\textbf{G1, attribution:} independently recorded agent and control
evidence lacks shared episode identity.
\textbf{G2, policy-at-time:} logs may omit the policy version,
threshold, or exception effective during the run.
\textbf{G3, privacy:} traces may retain raw prompts, credentials,
personal, or proprietary data when hashes or protected references suffice.
\textbf{G4, integrity:} evidence graphs may be mutated, replayed,
reordered, or poisoned~\cite{oraclepoisoning2026}, so the graph cannot
serve as its own integrity root.

% \paragraph{Scale and practical demands.}
% At scale, post-hoc assurance is a repeated join over action, control, and policy streams. Indexes reduce search, but each verdict still depends on identifier quality, event ordering, retained context, and later graph trust. ProofWeave instead standardises record creation, applies privacy transformation before persistence, assigns causal order at API receipt, anchors accepted records in an append-only ledger, and exposes a shared knowledge plane. A forensic query can then execute whenever a graph fragment is admitted rather than waiting for an incident. This changes forensics from retrospective reconstruction into continuous control testing.

% \vspace{-5pt}
\paragraph{Scale, practical demands, and limits of existing mechanisms.}
At scale, post-hoc assurance becomes a repeated join across action, control, and policy streams. Indexes can reduce search costs, but each verdict still depends on identifier quality, event ordering, retained context, and confidence in the reconstructed graph. As summarised in Table~\ref{tab:prior}, existing mechanisms address important aspects of this problem, but usually in isolation: operational logs and SIEM platforms support retention and search; provenance systems provide causal structure; agent-native representations capture agent activity; and graph reasoning enables semantic analysis. However, these approaches do not typically combine policy-at-time binding, independently attributable agent and control evidence, privacy minimisation before persistence, and an external integrity root within a shared cross-run evidence plane.

% ProofWeave addresses this combined requirement by standardising record creation, applying privacy transformations before persistence, assigning causal order at API receipt, anchoring accepted records in an append-only ledger, and exposing a shared knowledge plane. A forensic query can therefore execute whenever a graph fragment is admitted, rather than only after an incident has occurred. This reframes forensics from retrospective reconstruction as a form of continuous control testing. ProofWeave also replaces repeated cross-stream correlation with $O(k)$ indexed proof-path validation; Table~\ref{tab:numerical-cost} estimates reductions from $10{,}201$ to $1$ candidate binding, $10{,}201$ to $\approx 26$ validation operations, and from $0.79$ to $0.15$~MiB of evidence per project.
ProofWeave addresses this combined requirement by standardising record creation, applying privacy transformations, assigning causal order, anchoring accepted records in an append-only ledger, and exposing a shared knowledge plane. Forensic queries thus execute as evidence is admitted, reframing retrospective reconstruction into continuous control testing. Post-hoc correlation requires evaluating candidate bindings unverified potential matches between isolated action and control logs. ProofWeave replaces this ambiguity with $O(k)$ indexed proof-path validation; Table~\ref{tab:numerical-cost} estimates reductions from $10{,}201$ to $1$ candidate binding, $10{,}201$ to $\approx 26$ validation operations, and from $0.79$ to $0.15$~MiB of evidence per project.

\begin{table}[t]
\caption{Existing mechanisms and the remaining evidence-plane gap.}
\label{tab:prior}
\scriptsize
\begin{tabular}{@{}p{.20\columnwidth}p{.73\columnwidth}@{}}
\toprule
\textbf{Mechanism} & \textbf{Capability and limitation relative to \pw{}}\\
\midrule
Operational logs and SIEM & Retain and search events, but use producer-specific schemas and post-hoc joins; policy-at-time and causal semantics may be absent.\\
System provenance & Supports backward and forward causal analysis, but graph construction alone does not establish source attribution, privacy minimisation, or ledger-backed integrity~\cite{zipperle2022pids,li2021systemprovenance}.\\
Agent-native provenance & Represents model, tool, memory, and delegation activity, but does not necessarily bind independently attributable policy and control assertions through one evidence protocol~\cite{agentbom2026}.\\
Graph reasoning & Enables semantic analysis, but a poisoned graph may still drive internally consistent false conclusions~\cite{oraclepoisoning2026}.\\
Detector-specific files & Preserve findings from one tool, but do not form a cumulative cross-run evidence plane for testing the detector itself.\\
\bottomrule
\end{tabular}
\end{table}

\section{ProofWeave: One Evidence Chain}
\begin{figure}[t]
    \centering
    \includegraphics[width=0.8\linewidth]{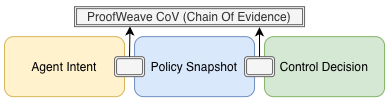}
    \caption{\small ProofWeave brings three attributes of evidence}
\label{fig:binding}
\end{figure}

\pw{} binds three artifacts that are typically stored separately: the Agent Intent and run context, the control action and decision, and the policy snapshot in force at the time of decision. Together, they form a single evidence record, so a later reviewer can determine what was attempted, what the control did, and which policy governed the outcome. Figure~\ref{fig:binding} presents this three-part view.

\subsection{Record-Time Binding}
For run $r$, let $I_r$ denote the Agent Intent and run context, $D_r$ the control action and decision, and $P_r$ the policy snapshot in force for that decision. The record-time binding API creates $E_r=\mathrm{Bind}(I_r,D_r,P_r).$
Binding writes the three record identifiers into the same signed evidence record while preserving the producer of each record. A shared \texttt{episode\_id}, repository, commit hash, parent event identifiers, policy identifier, sequence, and ledger position connect one run. RepoAudit traces, validator results, findings, errors, and completion events are attached to that episode.

Before append, the API checks the producer, event type, required identifiers, policy reference, payload hash, signature, and sequence. It removes or protects fields that are not needed for later checks. Raw prompts, full source files, credentials, and complete model responses stay outside the ledger and graph by default; the record keeps the identifiers, hashes, validator results, and encrypted references needed to repeat the check.

\subsection{Ledger, Graph, and Queries}
The binding API appends each accepted record to the append-only evidence ledger through \texttt{appendEvidence()}. Records are ordered and immutable; signatures, payload and predecessor hashes, sequence numbers, and episode identities expose tampering, reordering, or replay.
The Graph Weaver constructs the \pw{} evidence graph, $G_t=W(L_{\leq t})$, from accepted ledger records. Each node retains provenance to its source record. Model-suggested links remain marked as inferred and are validated before insertion; the graph cannot assert new recorded facts.
The Lineage Query API reconstructs the chain from intent and attempted action to the decision, governing policy snapshot, validators, and findings. The Policy Replay API re-evaluates the decision against its attached snapshot. The Detector Health API checks for a declared intent, identified policy, completed validators, and termination without an unresolved error.
Each API returns \texttt{pass}, \texttt{fail}, or \texttt{indeterminate}. Missing, rejected, unsigned, incomplete, or interrupted evidence produces \texttt{indeterminate}.

\begin{figure*}[!t]
    \centering
    \includegraphics[height=5cm]{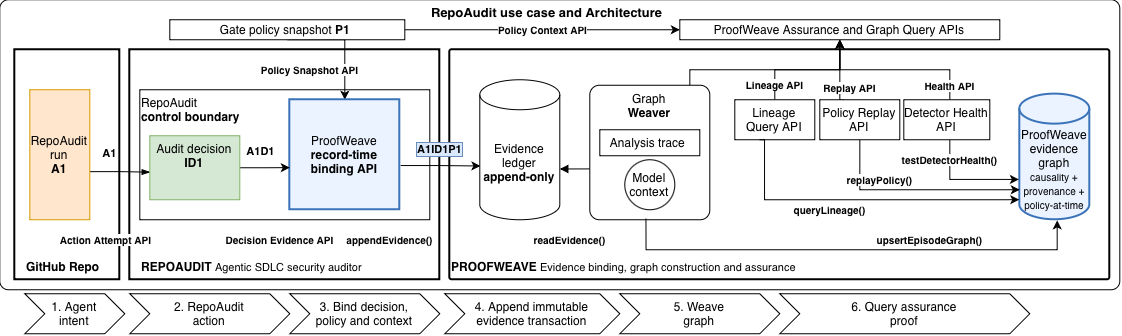}
    \caption{\small RepoAudit through the \pw{} architecture. }
    
    \label{fig:repoaudit-flow}
\end{figure*}

% \vspace{-5pt}
\section{Case Study: RepoAudit Through ProofWeave}
RepoAudit is an LLM-based repository code auditor. It selects audit starting points, follows inter-procedural paths, stores intermediate facts, runs validators, and emits findings~\cite{guo_repoaudit_2025}. The \pw{} integration does not change this method. It records how each RepoAudit run was carried out so that the run itself can be checked. Figure~\ref{fig:repoaudit-flow} uses the same terms as the implementation diagram.

% \vspace{-5pt}

% Export ProofWeave-RepoAudit-architecture-2.drawio as ProofWeave-RepoAudit-architecture.png before compiling.

% RepoAudit records the audit intent, action attempt, and audit decision. \pw{} binds the decision to the gate policy snapshot, appends the record to the evidence ledger, builds the evidence graph, and exposes lineage, policy replay, and detector health queries.

\subsection{RepoAudit Evidence Flow}
The case study follows the six steps in Figure~\ref{fig:repoaudit-flow}.
\begin{enumerate}
   
    \item \textbf{Agent intent.} Record the repository, commit, scope, control objective, model, prompt template, and run configuration.
    \item \textbf{RepoAudit action.} Record action attempt \texttt{A1}, analysis steps, validator calls, findings, errors, and completion state.
    \item \textbf{Bind decision, policy, and context.} Join audit decision \texttt{ID1} with action \texttt{A1}, gate policy snapshot \texttt{P1} at decision time.
    \item \textbf{Append the evidence record.} Apply the privacy checks and append signed \texttt{A1-ID1-P1} record to the evidence ledger.
    \item \textbf{Build the evidence graph.} Add Run, Decision, PolicySnapshot, Evidence, Finding, and ModelContext nodes with the recorded edge types shown in the diagram.
    \item \textbf{Query the evidence.} Use the lineage, policy replay, and detector health APIs to check how a finding was produced and whether the run completed the required work.
\end{enumerate}

\noindent During adoption, RepoAudit can keep its current file output and also call the \pw{} API. This dual-write adapter only sends records; it is not the evidence ledger. The append-only ledger is the store reached after the binding API accepts a record. Each run then adds one episode linked to the repository, commit, policy, model, prompt template, and control definition. This history supports checks for missing validators, incomplete runs, provider errors, stale evidence, and unexpected changes between runs.

\subsection{Checking RepoAudit Run Health}
The Detector Health API checks whether a RepoAudit run satisfies
the required operational evidence. For run $r$:
% \vspace{-1pt}
\[
\small
\begin{aligned}
T_{\mathrm{run}}(r) ={}&
\mathsf{Intent}(r)\land\mathsf{Policy}(r)
\land\mathsf{Complete}(r)\land\neg\mathsf{Error}(r)\\
&\land\ \forall f\in\mathsf{Findings}(r),\quad
V_{\mathrm{req}}(P_r,f)\subseteq V_{\mathrm{rec}}(f).
\end{aligned}
\]
Here, $V_{\mathrm{req}}$ and $V_{\mathrm{rec}}$ denote the required
and recorded validators, respectively. A run passes iff all
conditions hold. Policy conflicts, rejected findings, missing
validators, or blocked configurations produce \texttt{fail};
missing evidence or an interrupted run produces
\texttt{indeterminate}.
For comparable runs, the API compares coverage, validated findings, rejections, and errors against an allowed policy limit $\tau_P$. A larger change is flagged for review. The flag states that similar runs differed more than allowed; it does not, by itself, claim that RepoAudit is wrong.

\subsection{Prototype Checks, Limits, and Cost}

The prototype exercises a complete run, a missing validator event, a provider failure before completion, a finding inserted only into the graph, a changed or replayed ledger record, and an unexplained difference between comparable runs. Expected outcomes are pass, fail, indeterminate, rejected graph insertion, integrity failure, and a cross-run change flag.

Measurements include privacy filtering and API latency, ledger append, graph update, query latency, stored size, raw-to-protected evidence reduction, fault detection, and Graph Weaver model time. The ledger records which authenticated producer submitted each record and whether it was later modified. It does not prove that every submitted statement is true or that every event was reported. A separately protected collector could provide stronger coverage; this prototype evaluates only the API path in Figure~\ref{fig:repoaudit-flow}.

For protected event size $m$, ledger length $n$, graph additions $(v_r,e_r)$, and matched query size $k$, event checks cost $O(m)$, hash-chain append averages $O(1)$, optional Merkle updates cost $O(\log n)$, graph updates cost $O(v_r+e_r)$, and indexed queries cost $O(k)$ excluding index lookup. Model time is measured separately.

Table~\ref{tab:numerical-cost} shows that post-hoc correlation may produce up to 10,201 candidate bindings per verdict. ProofWeave instead validates a single record-time-bound path using approximately 26 lookup and integrity-checking steps. Privacy-minimised evidence also reduces estimated storage from $0.79$ to $0.15$ MiB per project.
\begin{table}[t]
\caption{Illustrative assurance cost for one RepoAudit project.}
\label{tab:numerical-cost}

\centering
\footnotesize
\setlength{\tabcolsep}{4pt}

\begin{tabular}{lcc}
\toprule
\textbf{Metric} &
\textbf{Post-hoc} &
\textbf{ProofWeave} \\
\midrule

Candidate bindings/verdict
    & $\leq 10{,}201$
    & $1$ \\

Validation operations/verdict
    & $\leq 10{,}201$
    & $\approx 26$ \\

Evidence storage/project
    & $\approx 0.79$ MiB
    & $\approx 0.15$ MiB \\

\bottomrule
\end{tabular}

% \vspace{2pt}
\raggedright
\scriptsize
Assumptions: 101 action records, 101 control records,
one policy snapshot, ledger length $n=10^6$, and matched query size $k=6$;
4-KiB raw events, a 1-KiB minimised transaction, and 0.5-KiB
derived graph state per transaction. All values are analytical estimates.
\end{table}

\balance
\bibliographystyle{ACM-Reference-Format}
\bibliography{ref}

\end{document}